\documentclass[a4paper,twocolumn,11pt,accepted=2023-09-18]{quantumarticle}
\pdfoutput=1
\usepackage[utf8]{inputenc}
\usepackage[english]{babel}
\usepackage[T1]{fontenc}
\usepackage[numbers]{natbib}
\usepackage{listings}

\usepackage{amsmath}
\usepackage{amssymb}
\usepackage[pagebackref=true]{hyperref}
\usepackage[resetlabels]{multibib}

\newcites{article}{References}
\usepackage{graphicx}% Include figure files
\usepackage{dcolumn}% Align table columns on decimal point
\usepackage{bm}% bold math
\usepackage{tikz}
\usepackage{hyperref}
\usepackage{orcidlink}
\usepackage{comment}

\usetikzlibrary{quantikz}
\begin{document}

%\preprint{APS/123-QED}

\title{Pulse-efficient quantum machine learning}% Force line breaks with \\

\author{André Melo}
\affiliation{Kavli Institute of Nanoscience, Delft University of Technology, P.O. Box 4056, 2600 GA Delft, The Netherlands}
\affiliation{IBM Quantum, IBM Netherlands,
Amsterdam, NH 1066 VH, The Netherlands}
\email{am@andremelo.org}
\author{Nathan Earnest-Noble \orcidlink{0000-0003-1441-9067}}
\affiliation{IBM Quantum, IBM T.\ J.\ Watson Research Center, Yorktown Heights, New York 10598, USA}
\author{Francesco Tacchino \orcidlink{0000-0003-2008-5956}}
\affiliation{IBM Quantum, IBM Research Europe -- Zurich, 8803 R\"{u}schlikon, Switzerland}
\email{fta@zurich.ibm.com}

\begin{abstract}
Quantum machine learning algorithms based on parameterized quantum circuits are promising candidates for near-term quantum advantage. Although these algorithms are compatible with the current generation of quantum processors, device noise limits their performance, for example by inducing an exponential flattening of loss landscapes. Error suppression schemes such as dynamical decoupling and Pauli twirling alleviate this issue by reducing noise at the hardware level. A recent addition to this toolbox of techniques is pulse-efficient transpilation, which reduces circuit schedule duration by exploiting hardware-native cross-resonance interaction. In this work, we investigate the impact of pulse-efficient circuits on near-term algorithms for quantum machine learning. We report results for two standard experiments: binary classification on a synthetic dataset with quantum neural networks and handwritten digit recognition with quantum kernel estimation. In both cases, we find that pulse-efficient transpilation vastly reduces average circuit durations and, as a result, significantly improves classification accuracy. We conclude by applying pulse-efficient transpilation to the Hamiltonian Variational Ansatz and show that it delays the onset of noise-induced barren plateaus.
\end{abstract}

\maketitle

\section{Introduction}
Quantum machine learning (QML) is a nascent area of research that has seen rapid developments over the last decade~\citearticle{wittek2014quantum, Biamonte2017, schuld2021machine,Mangini_2021}. Initial works in the field focused on developing quantum versions of existing classical algorithms, thereby achieving asymptotically faster runtimes~\citearticle{Wiebe2012, lloyd2013, Rebentrost2014}. However, these algorithms are beyond the capabilities of current quantum hardware, as they require the execution of rather deep quantum circuits and often rely on specific assumptions like quantum memories~\citearticle{PhysRevLett.100.160501} to overcome bottlenecks in data loading and readout. More recently, the fast technological progress and wide availability of noisy quantum processors~\citearticle{Preskill2018quantumcomputingin,Corcoles_2020} motivated the emergence of a second generation of QML algorithms based on parameterized quantum circuits (PQCs)~\citearticle{Benedetti_2019, cerezo2021variational, RevModPhys.94.015004,havlivcek2019supervised,PhysRevLett.122.040504,Tacchino2020IEEE,Abbas2021, Huang2021}. In this alternative QML paradigm, quantum computers may function as a co-processor working in tandem with classical computers. Because the circuit Ansätze can be shallow in depth, PQC-based algorithms are in principle compatible with the existing generation of quantum devices.

An important obstacle to the viability of QML in the near term is the presence of hardware noise~\citearticle{stilck2021limitations}. Coherent errors often simply shift the position of minima in the loss landscape, in which case they can be trained away~\citearticle{McClean2016}. In contrast, errors arising from incoherent noise have more adverse effects that hinder trainability and performance. As an example, incoherent errors cause the loss function of a large family of Ansätze to vanish exponentially with increasing number of layers, a phenomenon known as noise-induced barren plateaus (NIBP)~\citearticle{Wang2021}. A similar effect occurs in kernel-based methods where noise can lead to an exponential concentration of the kernel values~\citearticle{thanasilp2022exponential}.

A common approach to mitigate the effects of device noise is to use protocols that estimate improved expectation values through classical post-processing, a procedure known as error mitigation~\citearticle{Temme2017, Endo2018,PhysRevX.11.041036,PhysRevX.11.031057,Suchsland2021algorithmicerror,Cai2022}. While schemes such as zero-noise extrapolation~\citearticle{Temme2017, Endo2018} and virtual distillation~\citearticle{PhysRevX.11.041036} can significantly enhance the performances of noisy processors~\citearticle{kandala2019error}, they also introduce additional experimental or computational overhead and are can be ineffective in preventing noise-induced cost concentration~\citearticle{wang2021can, thanasilp2022exponential, https://doi.org/10.48550/arxiv.2210.11505,https://doi.org/10.48550/arxiv.2208.09385}. A complementary strategy is to apply error suppression schemes that suppress noise at the hardware level, such as dynamical decoupling~\citearticle{PhysRevLett.82.2417, PhysRevLett.121.220502} and Pauli twirling~\citearticle{cai2020mitigating}. A recently-introduced tool for hardware-level error suppression is pulse-efficient (PE) transpilation of cross-resonance gates~\citearticle{earnest2021pulse,PhysRevResearch.3.033171}. The core idea of this technique is to decompose two-qubit gates into hardware-native ones, such as echoed cross-resonance pulses on superconducting architectures based on fixed-frequency qubits. The echoes are then exposed to the transpiler to remove redundant single-qubit rotations. The resulting circuits often have significantly lower schedule durations, thereby mitigating errors introduced by finite coherence times. Crucially, PE transpilation requires no additional overhead or calibration. Refs.~\citearticle{earnest2021pulse, 9872062} applied this technique to combinatorial optimization tasks and demonstrated significant improvements over conventional CNOT-based transpilation. Related ideas were also explored in Refs.~\citearticle{meitei2021gate, liang2022variational, liang2023pan}, which directly optimized pulse parameters to address variational problems. However, a comprehensive study of the impact of PE transpilation on the performance of paradigmatic QML algorithms powered by parameterized quantum circuits was not available until now.

In this work, we demonstrate the application of PE transpilation to three paradigmatic QML tasks. To highlight the general applicability and versatility of our method, we conduct our experiments across three different IBM Quantum~\citearticle{IbmQuantum} backends using qiskit~\citearticle{Qiskit}. We begin by training Quantum Neural Networks (QNNs) on a synthetic dataset and observe that PE transpilation significantly improves the resulting training loss and classification accuracy. Afterwards, we apply PE transpilation to a quantum kernel circuit that we use to classify all 10 digits of the commonly-used MNIST dataset. When compared to CNOT-based transpilation, PE transpilation allows us to significantly extend the width of the kernel circuits and achieve a classification accuracy of $\approx 90\%$. Finally, we explicitly compute the effect of PE transpilation on NIBP. In particular, we study how the loss function of the Hamiltonian Variational Ansatz~\citearticle{kandala2017hardware} evolves for increasing number of layers and find that PE transpilation slows down the onset of the NIBP. 

\section{Application to quantum neural networks}
\begin{figure}[!tbh]
\centering
\sbox0{
\begin{tikzpicture}
\node[scale=0.95] {
\begin{quantikz}[column sep=5pt]
\lstick{$\ket{0}_0$} & \gate[wires=3]{\mathcal{U}_\text{FM}(\vec{x})} & \gate{R_Y(\theta_0)} & \ctrl{1} & \qw & \gate{R_y(\theta_{3})} & \meter{} \\
\lstick{$\ket{0}_1$} & & \gate{R_Y(\theta_1)} & \targ{} & \ctrl{1} & \gate{R_y(\theta_{4})} & \meter{} \\
\lstick{$\ket{0}_2$} & \qw & \gate{R_Y(\theta_2)} & \qw & \targ{} & \gate{R_y(\theta_{5})} & \meter{}
\end{quantikz}
};
\end{tikzpicture}
}%
\sbox1{
\begin{tikzpicture}
\node[scale=0.78] {
\begin{quantikz}[column sep=8pt]
 & \gate{H} & \gate{R_Z(2x_0)} & \ctrl{1}\gategroup[2,steps=3,style={dashed, rounded corners,fill=blue!20, inner xsep=0pt}, background]{{$R_{ZZ}$}} \qw \gategroup[2,steps=1,style={dashed, rounded corners,fill=red!15, inner ysep=-5pt, inner xsep=-2pt}, background] \qw & \qw & \ctrl{1} \qw & \qw & \qw & \qw & \qw \\
& \gate{H} & \gate{R_Z(2x_1)} & \targ{} & \gate{R_Z(x_{01})} & \targ{} & \ctrl{1} & \qw & \ctrl{1} & \qw \\
& \gate{H} & \gate{R_Z(2x_2)} \qw & \qw & \qw & \qw & \targ{} & \gate{R_Z(x_{12})} & \targ{} \qw & \qw
\end{quantikz}
};
\end{tikzpicture}
}%
\sbox2{\includegraphics[width=0.9\columnwidth]{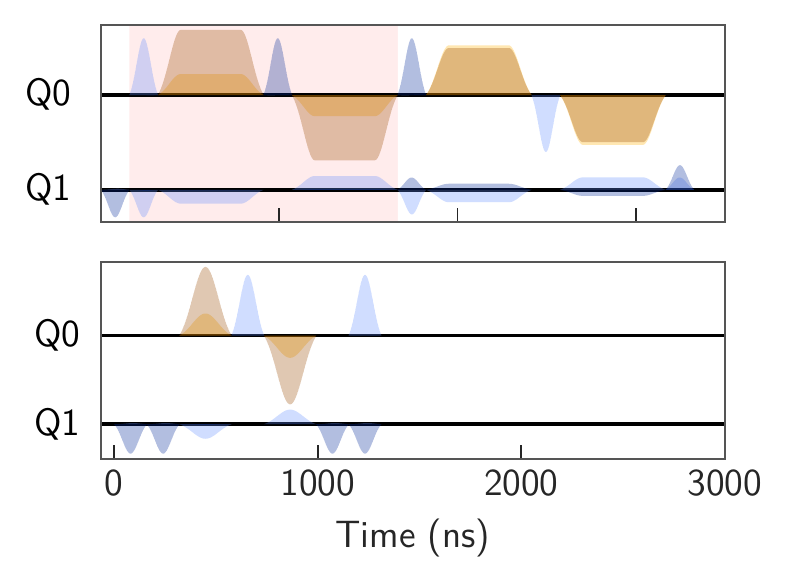}}%
\begin{tabular}{ll}
 (a) \\ \usebox0 \\
 (b) \\ \usebox1 \\
 (c) \\ \usebox2
\end{tabular}
\caption{(a) Quantum neural network architecture for binary classification. A forward pass of the network begins with an encoding stage that maps feature vectors $\vec{x}$ to a quantum state through a parameterized feature map $\mathcal{U}_\text{FM} (\vec{x})$. The second stage of the network consists of a variational form with parameterized $R_Y$ (whose angles are optimized with a classical routine) and CNOT gates applied on neighboring qubits. Finally, we measure the parity of the qubits in the $Z$ basis, with the fraction of even (odd) bitstrings representing the probability of class 0 (1). (b) A single layer of the parameterized feature map $\mathcal{U}_\text{FM} (\vec{x})$. Hadamard gates are applied on every qubit, followed by $R_{ZZ}$ rotations on every pair of neighboring qubits (highlighted in blue). (c) Pulse schedules that implement an $R_{ZZ}(0.5)$ gate on \emph{ibmq\_guadalupe} through a conventional CNOT-based approach (top panel) and pulse-efficient transpilation (bottom panel). The section highlighted in red in the top panel corresponds to the pulses that implement a single CNOT gate through cross-resonance. Pulse-efficient transpilation significantly decreases the schedule duration resulting in higher circuit fidelities.}
\label{fig:qnn_circuit}
\end{figure}
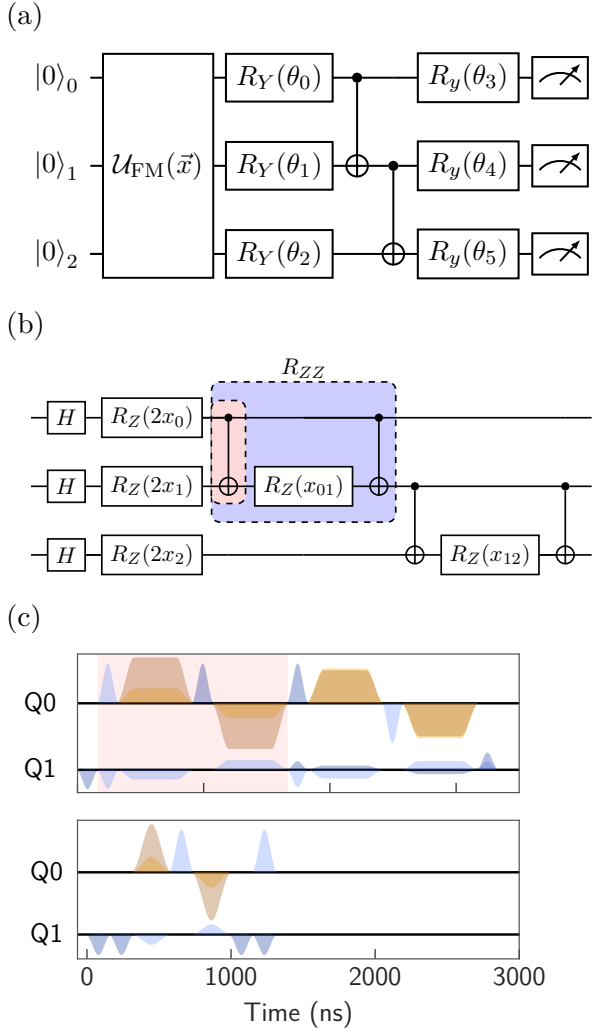

\begin{figure*}[!tbh]
    \centering
    \includegraphics{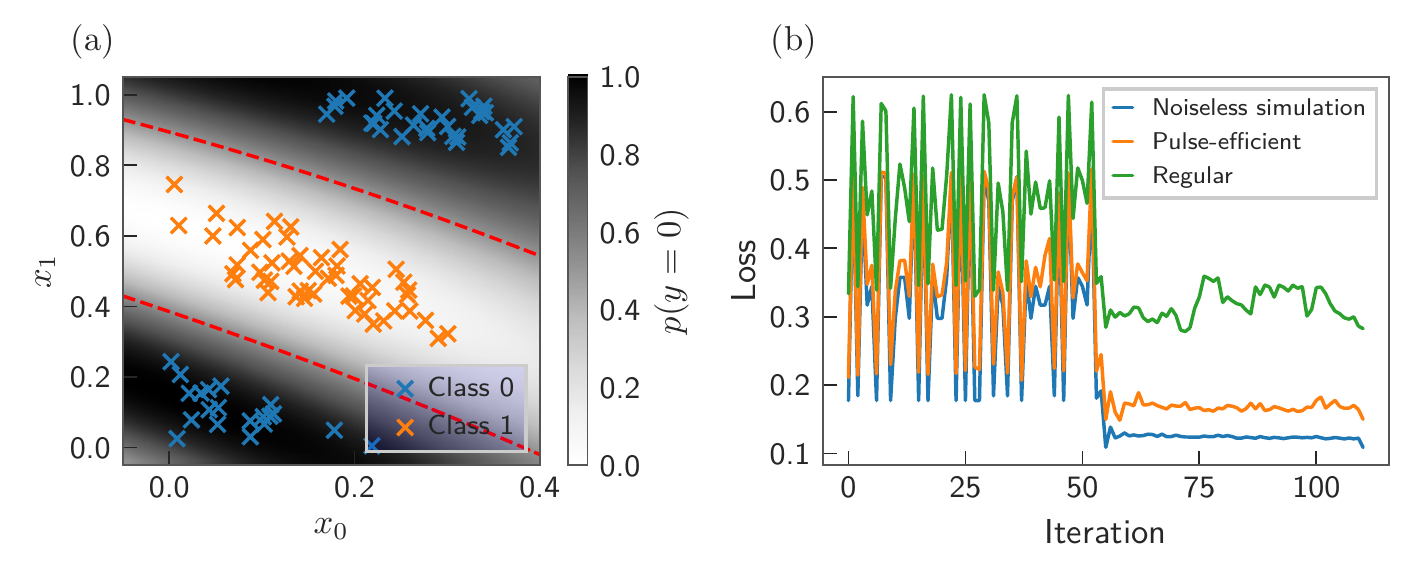}
    \caption{Training a quantum neural network on \emph{ibmq\_jakarta}. (a) Example of a synthetic two-class dataset generated from the QNN shown in Fig.~\ref{fig:qnn_circuit} with $n=2$ features. The heat map represents the probability assigned to class 0 by the QNN with the weights set to those used to generate the binary dataset. The red dashed lines correspond to the decision boundaries $p(y=0) = p(y=1) = 0.5$ of the model. (b) Convergence of the training loss on $n=4$ qubits dataset after 120 iterations of the SPSA algorithm. The blue, orange and green curves show the training loss of the simulated, pulse-efficient, and regular quantum neural networks, respectively.}
    \label{fig:dataset_training}
\end{figure*}

\begin{figure}[!tbh]
    \centering
    \includegraphics{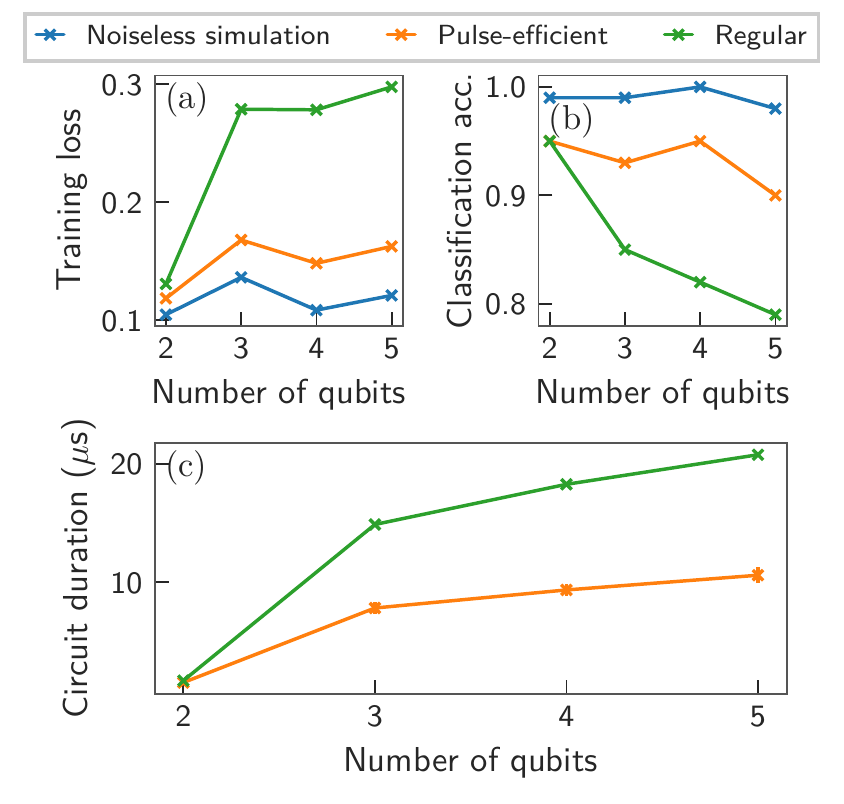}
    \caption{Comparison of the performance of quantum neural networks with pulse-efficient and regular transpilation. (a) Training loss after 120 iterations of the SPSA algorithm. (b) Classification accuracy on the test set. (c) Average schedule duration of QNN circuit. The pulse-efficient circuits have significantly lower schedule duration, which improves circuit fidelity and classification accuracy.}
    \label{fig:qnn}
\end{figure}

QNNs are one of the leading variational algorithms for QML~\citearticle{arxiv.1802.06002, PhysRevA.101.032308, Benedetti_2019,liu2022representation,liu2023analytic}. A typical QNN architecture for classification tasks is comprised of three main steps. First, classical data $\vec{x}\in \mathbb{R}^n$ is encoded onto a quantum state through a parameterized feature map $\mathcal{U}_\text{FM} (\vec{x})$. The resulting state $\mathcal{U}_\text{FM} (\vec{x}) |0 \rangle^{\otimes n}$ is then fed to a variational ansatz $\mathcal{U}_\text{v}(\vec{\theta})$, where $\vec{\theta}$ denotes a set of trainable parameters. Finally, the expectation value of an observable $\mathcal{O}$ is measured and classically post-processed to yield the model predictions $y_p = f(\vec{x} | \vec{\theta})$ and the loss function $L(f(\vec{x} | \vec{\theta}), \vec{y})$, where $y$ are the true labels. A classical optimization algorithm then iteratively varies $\vec{\theta}$ to minimize $L$.

We begin by studying the impact of PE transpilation on a binary classification task with the QNN shown in Fig.~\ref{fig:qnn_circuit} (see App.~\ref{app:pe} for a description of the underlying mechanism which enables our specific PE method). We consider an architecture similar to the one reported in Ref.~\citearticle{Abbas2021} but restrict entangling operations to neighboring pairs of qubits in order to avoid prohibitively large circuit depths. A forward pass of the QNN starts with two layers of the feature map proposed in Ref.~\citearticle{havlivcek2019supervised} (Fig.~\ref{fig:qnn_circuit} (b)). A single layer applies Hadamard and $R_Z$ gates on all qubits, where the $R_Z$ rotation angle is related to the feature values by the relation $2x_i$. This is followed by $R_{ZZ}$ operations on neighboring pairs of qubits, where the rotation angle $x_{ij} = 2 (\pi - x_i)(\pi - x_j)$ depends on products of the features. We then apply a variational ansatz that consists of parameterized $R_Y$ gates applied on every qubit, followed by CNOTs on neighboring qubits, and a final set of parameterized $R_Y$ (Fig.~\ref{fig:qnn_circuit} (a)). The angles $\theta_i$ of the $R_Y$ operations are the training parameters that are optimized classically to fit a given target function. Finally, we measure the parity of the output bit strings
\begin{equation}
    m(\vec{x}, \vec{\theta}) = |\langle 0 | \mathcal{U}_\text{FM}^\dag(\vec{x}) \mathcal{U}_\text{v}^\dag(\vec{\theta}) P \mathcal{U}_\text{v}(\vec{\theta}) \mathcal{U}_\text{FM}(\vec{x}) | 0 \rangle|^2
\end{equation}
where $P = \prod_i Z_i$ is the parity operator and $Z_i$ is the standard Pauli $Z$ operator acting on the $i$-th qubit.
We associate even parity with class 0 and odd parity with class 1. 

We benchmark the QNN performance on a synthetic two-class dataset with standard and PE transpilation. To ensure the QNN can distinguish between the two classes with high accuracy, the dataset is generated via the QNN model by fixing the trainable parameters $\vec{\theta_s}$ in such a way that the separation of the classes is maximised. The training procedure is then carried out starting from a new, random initialisation of the parameters, and should ideally recover the set used to generate the data. More formally, we uniformly sample $600$ feature vectors $\vec{x} \in [0, 1)^n$ and compute their parity $m(\vec{x}, \vec{\theta})$ through noiseless simulations. Using the L-BFGS-B optimizer~\citearticle{zhu1997algorithm}, we search for QNN parameters $\vec{\theta_s}$ that maximize the average absolute parity $\frac{1}{600} \sum_i |m(\vec{x_i}, \vec{\theta_s})|$. Out of this set of feature vectors, we further select the 50 samples for each class with the largest absolute parity expectation value.
Fig.~\ref{fig:dataset_training}(a) shows an example of the resulting dataset for 2 qubits alongside the probability of observing class 0 $p(y=0, \vec{x}, \vec{\theta_s})$. We observe that the decision boundary of the QNN correctly separates the two classes.

We train the QNNs on \emph{ibmq\_jakarta} using a cross-entropy loss function and 50 iterations of Spall's SPSA stochastic gradient descent algorithm~\citearticle{spall2000adaptive} with an automated calibration phase~\citearticle{kandala2017hardware} of 50 iterations. Moreover, we apply readout error mitigation~\citearticle{PhysRevA.103.042605} and use 100 samples both for training and testing the networks. We show example training curves for $n = 3$ qubits in Fig.~\ref{fig:dataset_training}(b) which converge almost immediately after the initial calibration stage. In Fig.~\ref{fig:qnn}(a-b), we compare the training loss and testing classification accuracy of pulse-efficient and regular QNNs with $n=2$ to $5$ qubits. While the performance of the standard QNN deteriorates rapidly after $n=2$, the PE QNN remains competitive with the performance of the noiseless simulation throughout the whole range of $n$. We attribute this improvement to a reduction in incoherent error due to the shorter schedule duration in PE circuits (Fig.~\ref{fig:qnn}(c)). More specifically, as we show in Fig.~\ref{fig:qnn_circuit}(c) PE transpilation significantly shortens the duration of $R_{ZZ}$ gates~\citearticle{earnest2021pulse} and hence of the feature map portion of the circuit.

\section{Application to quantum kernels}
\label{sec:kernel}
\begin{figure*}[!tbh]
    \centering
    \includegraphics{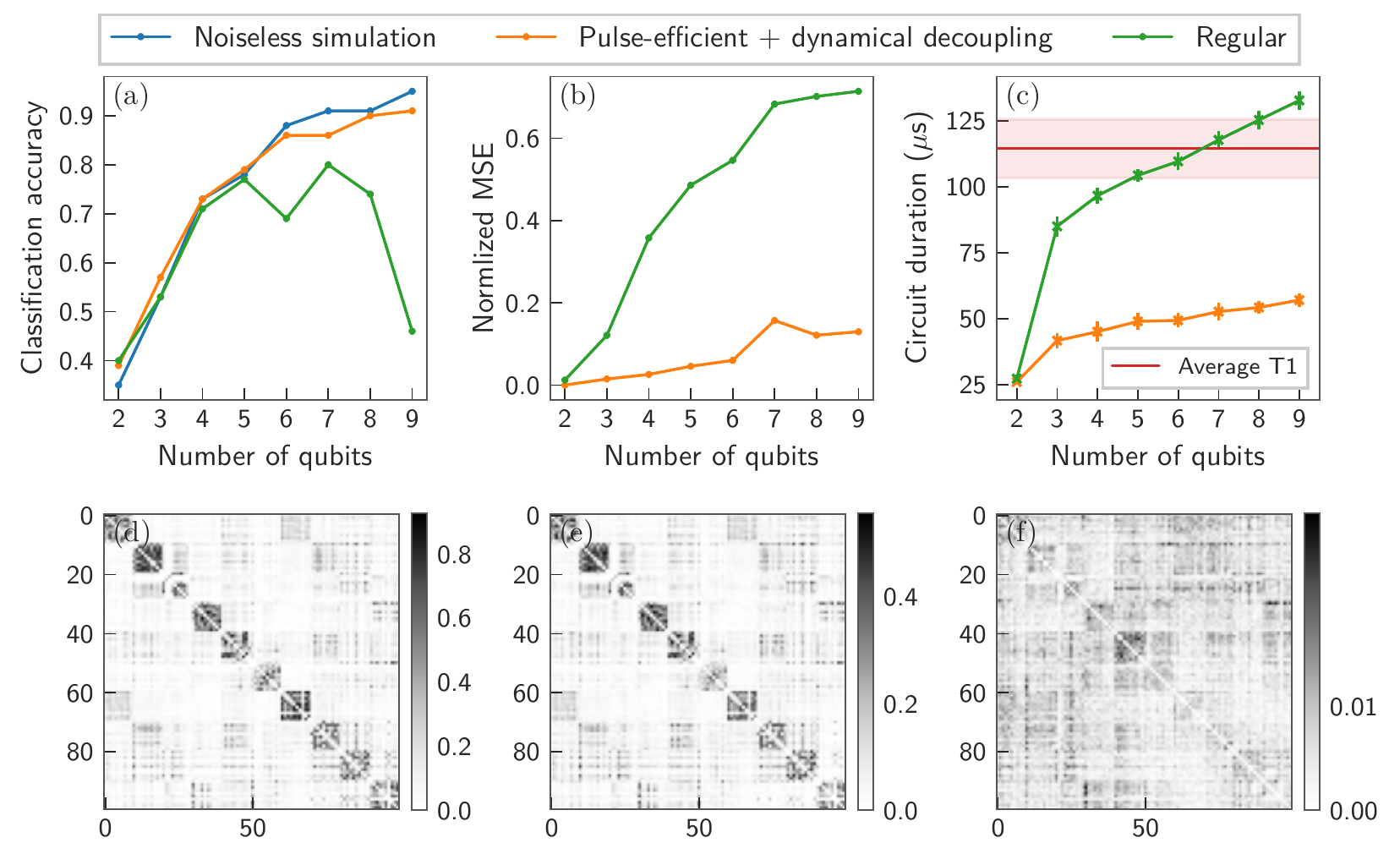}
    \caption{Impact of pulse-efficient transpilation on quantum kernel classification run on \emph{ibmq\_montreal}. (a) Classification accuracy on test dataset for a noiseless simulation, pulse-efficient, and regular circuits. (b) Normalized mean square error of the training kernel matrix compared to the simulated kernel matrix. (c) Average circuit duration for pulse-efficient and regular circuits compared to the $T_1$. In the bottom panel we show the training kernel matrices at $n=9$ qubits obtained with (d) noiseless simulations, (e) pulse-efficient circuits, and (f) regular circuits.}
    \label{fig:kernel}
\end{figure*}

We now turn to investigating the impact of PE transpilation on fidelity-based quantum kernel classification. This class of algorithms uses a quantum feature map to compute a similarity measure between input data points~\citearticle{PhysRevLett.122.040504, havlivcek2019supervised, wilson2018quantum, Huang2021, schuld2021supervised}
\begin{equation}
    K(\vec{x_i}, \vec{x_j}) = |\langle 0 | \mathcal{U}_\text{FM}^\dag(\vec{x_i}) \mathcal{U}_\text{FM}(\vec{x_j}) | 0 \rangle|^2.
\end{equation}
The resulting Gram matrix $K$ is then fed to a classical kernel method such as a support vector machine~\citearticle{zbMATH00810144} to predict the corresponding labels~$\vec{y}$.

For this experiment we use the same feature map as in the QNN case presented above, but increase the depth to 4 in order to achieve higher classification accuracy. Following the approach outlined in Ref.~\citearticle{havlivcek2019supervised}, we estimate the feature vector kernel function for all pairs of training data $\vec{x_i}, \vec{x_j}$ using 8192 shots. Specifically, we apply the circuit $\mathcal{U}_\text{FM}^\dag(\vec{x_i}) \mathcal{U}_\text{FM}^\dag(\vec{x_j}) |0\rangle^{\otimes n}$ and then measure all qubits in the $Z$ basis. The kernel entry $K(\vec{x_i}, \vec{x_j})$ then corresponds to the frequency of the zero bitstring $0^n$.
Having repeated this process for all the training data, we feed the resulting kernel matrix to a conventional support vector machine implemented with \emph{scikit-learn}~\citearticle{scikit-learn}. For this classification task, we choose the MNIST dataset, a popular real-world database of handwritten digits~\citearticle{Dua2019}, and use 10 training and testing samples for each of the ten digits. The kernel chosen for this experiment uses a the number of input features equal to the number of qubits. In the case of the MNIST dataset, the resolution of the images exceeds the number of qubits we use. We therefore reduce the number of features through a truncated singular value decomposition, a standard dimensionality reduction procedure.

Due to the sparsity of the kernel circuits, the qubits experience large idle times that lead to error accumulation~\citearticle{https://doi.org/10.48550/arxiv.2105.03406}. To mitigate this source of noise, we combine PE transpilation with a dynamical decoupling protocol. Whenever qubit $i$ has an idle time $T_\text{idle}$ larger than twice the single qubit gate time, we apply a dynamical decoupling sequence $\tau/2 - X_p - \tau - X_m - \tau/2$ with delay times $\tau = T_\text{idle} - 2T_{X_{p/m}}$ where $X_{p/m}$ are positive/negative $\pi$ pulses around the $x$-axis and $T_{X_{p/m}}$ their duration. 

We run the kernel circuits on linearly connected subsets of qubits on \emph{ibmq\_montreal} with PE and regular transpilation, along with noiseless simulations. Fig.~\ref{fig:kernel}(a) shows the testing classification accuracy as a function of the number of qubits for all three methods. Focusing first on the simulated curve, the classification accuracy increases monotonically with the number of qubits. This occurs because the number of qubits increases concomitantly with the number of training features, which makes it easier to distinguish different digits. Turning to the device runs, the performance of the regular circuit remains very close to the ideal curve up to 5 qubits, after which it degrades rapidly and stays below 80\%. This sharp turning point coincides with the average circuit duration becoming comparable with the average device $T_1$, see Fig~\ref{fig:kernel}(c). In contrast, the PE transpilation circuit durations are always well below the coherence limit of the device, thereby yielding classification accuracies that reach 90\% and closely track the simulated values. To further quantify the performance of the device runs, in Fig~\ref{fig:kernel}(c) we show the normalized mean square error of the experimental kernel matrices compared to the simulated matrix $K^\mathrm{sim}$:
\begin{equation}
\mathrm{NMSE} (K) = \frac{\sum_{ij}(K^\mathrm{sim}_{ij} - K_{ij})^2}{\sum_{ij}( K^\mathrm{sim}_{ij})^2}.
\end{equation}
The error curves show a similar trend to the average circuit durations, with the error of the regular circuits increasing much faster than the PE circuits. In Fig.~\ref{fig:kernel}(d-f) we show training kernel matrices with $n=9$ qubits for all three methods. The PE transpilation kernel matrix is close to the simulated matrix and has an approximately block-diagonal structure, indicating the feature map is capable of separating the digits with high accuracy. In contrast, the regular transpilation matrix is mostly devoid of structure, which results in significantly lower classification accuracy. Moreover, its matrix elements are close to 0, signaling that the underlying bitstring distribution is extremely noisy. 

In App.~\ref{app:dd} we perform additional experiments to estimate how much of the performance boost we observe can be attributed to PE transpilation versus dynamical decoupling and find that PE transpilation is the main driver of the improvement.

\section{Impact on noise-induced barren plateaus}
\begin{figure}[!tbh]
    \centering
    \includegraphics{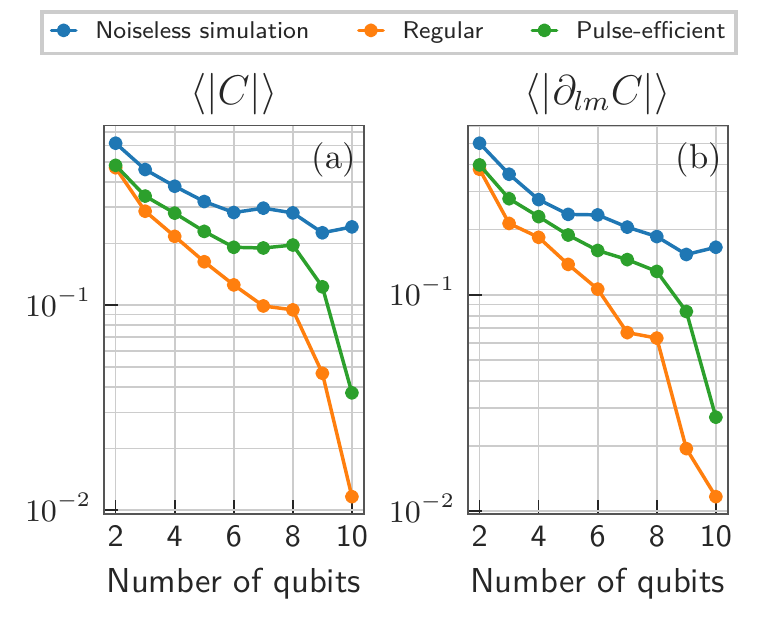}
    \caption{Mitigation of noise-induced barren plateaus through pulse-efficient transpilation. We consider an ansatz similar to the Hardware Variational Ansatz with a number of layers that increases linearly with the number of qubits. By sampling over 100 sets of random parameters we compute the average (a) loss function, and (b) partial derivative with respect to the last $\beta_{i, i+1}$ parameter.}
    \label{fig:nibp}
\end{figure}

In our last experiment, we investigate the impact of PE transpilation on NIBP. We implement the Hamiltonian Variational Ansatz for the transverse field Ising model as considered in Refs.~\citearticle{PRXQuantum.1.020319} and ~\citearticle{Wang2021}. We consider a linearly connected chain of spins, such that a single layer of the ansatz is given by
\begin{equation}
    U_\mathrm{TFIM} = \exp \left(-i H_{x} \right) \exp \left(-i H_{zz} \right),
\end{equation}
where
\begin{align}
    H_{x} &= \sum_{i=0}^{N-1} \gamma_{i} \sigma_i^x, \\
    H_{zz} &= \sum_{i=0}^{N-2} \beta_{i, i+1} \sigma_i^z \sigma_{i+1}^z,
\end{align}
and $\sigma_\alpha^i$ are the conventional Pauli matrices acting on the $i$-th qubit. To study the onset of NIBP, we study the behavior of a local observable with increasing number of qubits $n$ and ansatz layers $L$. Following~\citearticle{Wang2021}, we measure the local parity of the first two qubits $O = Z_0 Z_1$ along with its derivative with respect to the last $\beta_{i, i+1}$ parameter. Further, we set the number of layers to increase linearly with the number of qubits $L = 2(n-1)$ and perform our runs on \emph{ibmq\_montreal}.

In Figure~\ref{fig:nibp} we show the cost function and its partial derivative with respect to the last $\beta_{i, i+1}$ parameter averaged over 100 random parameter sets. In the case of noiseless simulations, both curves appear to slowly decay polynomially with increasing number of qubits. On the other hand, the device runs show a noticeable exponential decay starting at around $n=8$ qubits, which we attribute to the onset of NIBP. However, both the average loss function and derivative of the PE transpilation circuits are consistently above those of the regular circuits. Although PE transpilation does not remove NIBP, it has more favorable scaling which would allow going to higher depths of number of qubits when compared with regular transpilation.

\section{Discussion}
In this work, we studied the impact of PE transpilation on the performance of near-term QML algorithms. We began by performing binary classification of a synthetic dataset with a QNN, where we found that PE circuits achieved significantly higher classification accuracy and lower training loss. Secondly, we used quantum kernel estimation to classify a real-world dataset of handwritten digits. Combining PE transpilation with dynamical decoupling allowed us to accurately estimate kernels up to 9 qubits and achieve 90\% classification accuracy, whilst regular circuits remain below 80\%. Lastly, we studied the onset of NIBP on a commonly-used ansatz for quantum chemistry. Our results show that PE transpilation slows down the onset of NIBP, which allows executing variational quantum algorithms at higher numbers of qubits when compared with regular transpilation. 

Our results highlight a key advantage of PE transpilation, namely that it introduces no additional overhead or calibrations and is compatible with most qiskit-based programs with minimal modifications. Furthermore, we observe that it consistently improves circuit performance across different models and devices. This makes our proposed approach particularly appealing for applications and use cases that rely on remote device access and control. We expect that these improvements extend to most protocols featuring parameterized $R_{ZX}(\theta)$ gates: these natively appear in a broad class of quantum algorithms, such as Hamiltonian simulation schemes~\citearticle{Miessen2023,kim2023scalable} -- with potential applications to optimization~\citearticle{Weidenfeller2022} and sampling problems~\citearticle{Mazzola2021,Layden2022} -- and unitary coupled cluster circuits in quantum chemistry.

While finalizing this work, we became aware of a recent preprint~\citearticle{https://doi.org/10.48550/arxiv.2211.00350} that also applies PE transpilation to PQCs designed for quantum chemistry and optimization tasks.

\section*{Acknowledgements}
We are thankful to Caroline Tornow, Daniel J.~Egger and Kunal Sharma for useful discussions. We acknowledge the use of IBM Quantum services for this work. IBM, the IBM logo, and ibm.com are trademarks of International Business Machines Corp., registered in many jurisdictions worldwide. Other product and service names might be trademarks of IBM or other companies. The current list of IBM trademarks is available at \url{https://www.ibm.com/legal/copytrade}.

%\bibliography{references}
\bibliographystylearticle{quantum}
\bibliographyarticle{references}
%\bibliographystyleweb{quantum}
%\bibliographyweb{references}

\appendix
\begin{figure*}[!tbh]
    \centering
    \includegraphics{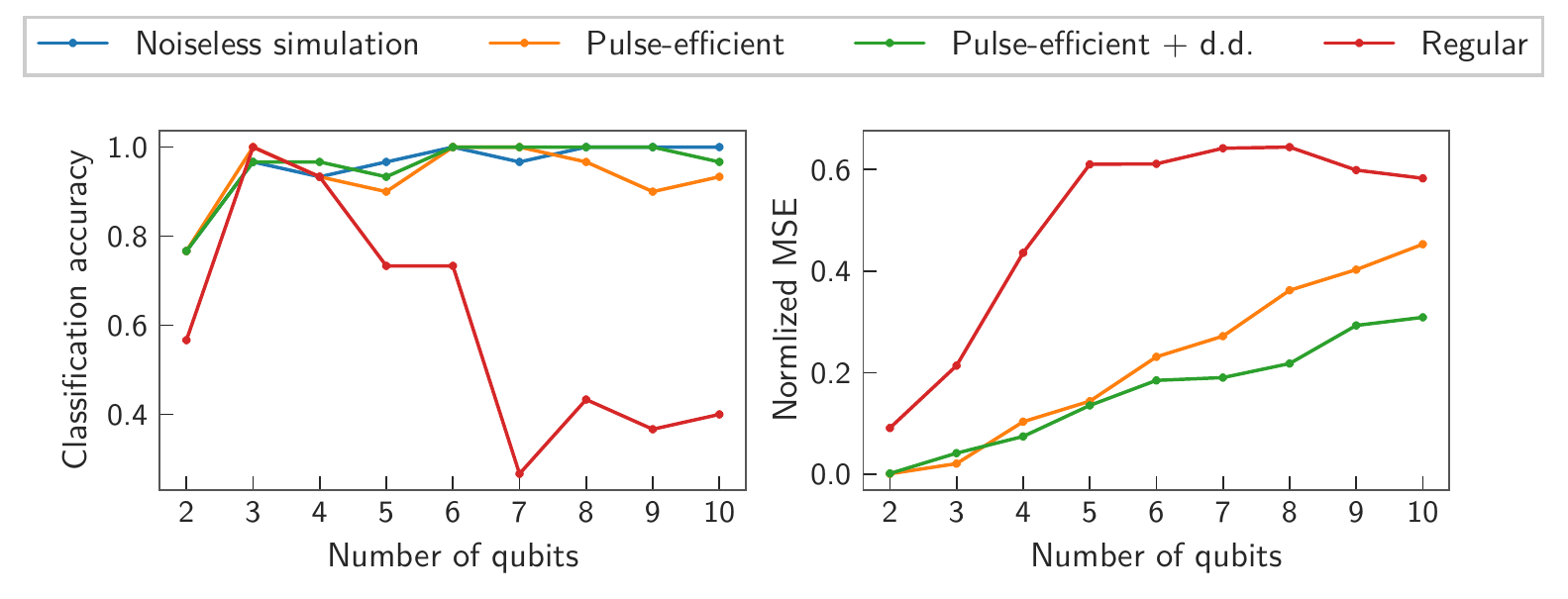}
    \caption{Impact of pulse-efficient transpilation on quantum kernel classification run on \emph{ibmq\_gaudalupe}.}
    \label{fig:kernel_appendix}
\end{figure*}

\section{Pulse-efficient cross-resonance circuits}
\label{app:pe}
In this appendix we briefly review pulse-efficient transpilation for cross-resonance-based hardware (typically coupled fixed frequency transmons).
The cross-resonance interaction arises by driving a control qubit at the target qubit's frequency.
Within the two-level approximation, the resulting time-independent Hamiltonian reads
\begin{equation}
    H_{CR} = \frac{1}{2} \left (Z \otimes B + I \otimes C  \right),
\end{equation}
where B = $\omega_{ZI} I + \omega_{ZX} X + \omega_{ZY} Y$, C = $\omega_{IX} X + \omega_{IY} Y + \omega_{IZ} Z$, and $I, X, Y, Z$ are Pauli matrices~\citearticle{chow2011simple, magesan2020effective, malekakhlagh2020first}.
By using echoed cross-resonance pulses with rotary tones \citearticle{sheldon2016procedure, sundaresan2020reducing}, it is possible to isolate the $ZX$ interaction and thus implement the unitary $R_{ZX} (\theta) = \exp \{-i ZX \theta / 2 \}$ with good accuracy.
To first approximation, the rotation angle of this conditional rotation is given by $\theta = t_{CR} \omega_{ZX}(A)$, where $\omega_{ZX}(A)$ is a non-linear interaction term that depends on the amplitude of the cross-resonance pulse.
IBM Quantum backends leverage the cross-resonance interaction to implement CNOT gates constructed with echoed rotations $R_{ZX}(\pi/2) = \mathrm{CR}(\pi/4) X \mathrm{CR}(-\pi/4) X$.
Here, $\mathrm{CR}$ are the non-echoed cross-resonance pulses, typically shaped as flat-top gaussians.
Together with a complete set of one-qubit gates, this CNOT gate is then used as a primitive to synthesize arbitrary two-qubit gates.
However, it is possible to implement arbitrary rotations $R_{ZX}(\theta)$ by appropriately scaling the cross-resonance pulses~\citearticle{earnest2021pulse}.
The core idea of PE transpilation is to leverage this native parametric gate to decrease circuit duration and achieve higher fidelities.
The scheme works as follows.
Using Cartan's decomposition, we first rewrite a CNOT-transpiled circuit in terms of parameterized, non-echoed $R_{ZX}(\theta)$ rotations.
Then, we expand the $R_{ZX}(\theta)$ gates and expose its echoed implementation $\mathrm{CR}(\theta/2) X \mathrm{CR}(-\theta/2) X$.
A final transpilation pass removes redundant single-qubit rotations, leaving at most one single-qubit rotation between non-echoed $R_{ZX}$ pulses. 
Though non-linear behaviour can result in coherent over or under rotations, this method achieves significant reductions in circuit duration for certain gates (such as the $R_{ZZ}(\theta)$ interaction used in this work) compared to conventional CNOT-based transpilation.

\section{Pulse-efficient kernel classification without dynamical decoupling}
\label{app:dd}

In Section~\ref{sec:kernel}, we showed kernel estimation results for circuits with PE transpilation and dynamical decoupling. A natural follow-up question is how much of the performance improvement can be attributed to each of the two error suppression strategies. To address this question, we run a smaller set of experiments on \emph{ibmq\_gaudalupe} to classify digits 0, 7, and 9. We execute the kernels with PE transpilation with and without dynamical decoupling and show the resulting classification accuracy and NMSE in Fig.~\ref{fig:kernel_appendix}. Although dynamical decoupling has a sizeable effect, we conclude that PE transpilation is the primary driver of the performance improvement over regular circuits. 

\end{document}